\documentclass{article}

\usepackage{amsmath, amsthm, amscd, amsfonts, amssymb, graphicx, color}
\usepackage[bookmarksnumbered, colorlinks, plainpages]{hyperref}
\usepackage{pspicture,pst-all,multido}
\theoremstyle{definition}

\theoremstyle{remark}

\numberwithin{equation}{section}

\newcommand{\C}{\mathbb{C}}
\newcommand{\R}{\mathbb{R}}

\newcommand{\norm}[1]{\parallel\!\!#1\!\!\parallel}
\renewcommand{\d}{\operatorname{d}}
\newcommand{\e}{\operatorname{e}}
\renewcommand{\i}{\operatorname{i}}

\newcounter{envcount}%

\newenvironment{AC}%
{\vspace{\bigskipamount}\refstepcounter{envcount}\textbf{(\theenvcount)\enspace Asymptotic causality.}}%
  {\vspace{\bigskipamount}}

\newenvironment{Def}%
{\vspace{\bigskipamount}\refstepcounter{envcount}\textbf{(\theenvcount)\enspace Definition.}}%
  {\vspace{\bigskipamount}}

\newenvironment{TPP}%
{\vspace{\bigskipamount}\refstepcounter{envcount}\textbf{(\theenvcount)\enspace  }}%
  {\vspace{\bigskipamount}}

\newenvironment{The}%
{\vspace{\bigskipamount}\refstepcounter{envcount}\textbf{(\theenvcount)\enspace Theorem.}\itshape}%
  {\vspace{\bigskipamount}\upshape}

{\vspace{\bigskipamount}\refstepcounter{envcount}\textbf{(\theenvcount)\enspace Theorem}\itshape}%
  {\vspace{\bigskipamount}\upshape}

{\vspace{\bigskipamount}\refstepcounter{envcount}\textbf{(\theenvcount)\enspace Proposition.}\itshape}%
  {\vspace{\bigskipamount}\upshape}

\newenvironment{Cor}%
{\vspace{\bigskipamount}\refstepcounter{envcount}\textbf{(\theenvcount)\enspace Corollary.}\itshape}%
  {\vspace{\bigskipamount}\upshape}

\newenvironment{Lem}%
{\vspace{\bigskipamount}\refstepcounter{envcount}\textbf{(\theenvcount)\enspace Lemma.}\itshape}%
  {\vspace{\bigskipamount}\upshape}

{\vspace{\bigskipamount}\refstepcounter{envcount}\textbf{(\theenvcount)\enspace}}%
  {\vspace{\bigskipamount}}

\theoremstyle{definition}
\swapnumbers

\begin{document}
\setcounter{page}{1}
\pagenumbering{arabic}

\begin{center}
{\Large Rebound Motion of  Localized  Dirac Wavefunctions}\\  

\vspace{1cm}
Domenico P.L. Castrigiano\\
Technischen Universit\"at M\"unchen, Fakult\"at f\"ur Mathematik, M\"unchen, Germany\\

\smallskip

{\it E-mail address}: {\tt
castrig\,\textrm{@}\,ma.tum.de
}
\end{center}

\begin{quote}  It is shown that  the carrier  of a bounded localized free Dirac wavefunction  shrinks from infinity and subsequently expands  to infinity again. The motion occurs isotropicly at the speed of light. In between there is the phase of rebound, which is limited in time and space in the order of the diameter of the carrier at its minimal extension. This motion proceeds anisotropicly and abruptly as for every direction in space there is a specific time, at which the change  from shrinking to expanding happens instantaneously.  Asymptotically, regarding the past and the future as well, the probability of position concentrates up to $1$   within any spherical shell whose outer radius increases at light speed.

\end{quote}

\section{Introduction}   
The present investigations provide new insights  regarding particle localization with causal time evolution. This  concept in relativistic quantum theory  is the matter of incessant research and up to now has not reached a commonly accepted resolution. For  the by now vast literature on the subject see \cite{CL15}, \cite{C18} and  the references therein as e.g.\,\cite{LCE} for more recent contributions. It is known for a long time that localization in the sense of Wightman (WL) \cite{W62}  is not compatible with causality if the energy operator $H$ is semi-bounded.  As shown by Schlieder \cite{Sch71}, relying on a theorem by Borchers, causality and semi-boundedness of energy  imply confinement. This result has been generalized notably by Hegerfeldt \cite{H01} reducing the premises and simplifying the proof. One notes that  Hegerfeldt's work  provided new impetus to the research in various directions. Hence causality and localizability may be reconcilable  only if $H$ is unbounded above and below, and  the  challenge of unbounded negative energies arises.\\
\hspace*{6mm}
Actually, causality of time evolution determines the Hamiltonian $H$ for  a massive system with finite spinor dimension
rather definitively.  According to   \cite{CL15}, for every positive mass, there is a sequence of Dirac tensor-localizations \cite[Eq.\,(2)]{CL15}, which constitute   a complete set of inequivalent irreducible WL with causal time evolution. They follow from Dirac's system  enlarging the spinor space by a simple tensor-construction. Therefore up to unitary equivalence, without assuming relativistic symmetry  one ends up  with  a finite orthogonal sum $H=\bigoplus_m\,\nu_m\; H_m$
for positive  masses $m$,  finite multiplicities  $\nu_m$, and  $H_m$ the Dirac operator for mass $m>0$ at the right hand side of (\ref{DOPR}). The states of the system are given by the normalized wavefunctions $\psi$, where $|\psi(x)|^2$ is  the position probability density.
\\
\hspace*{6mm}
This result shows that the Dirac system is fundamental. For studying the localization of a massive particle  with causal time evolution  it suffices henceforth to deal  with the latter.
\\
\hspace*{6mm}
The crux are the bounded localized sates, i.e. the normalized wavefunctions $\psi$ with bounded carrier, which necessarily are  a superposition of a non-vanishing positive and negative energy component. Conversely the carriers  of the normalized purely positive energy wavefunctions $\Psi$, which represent the Dirac electron states,     are not bounded but essentially dense in  $\R^3$   \cite[Cor.\,1.7]{T92},  \cite[(7) Theorem, (80) Cor.]{C18}, which roughly speaking  means  that the electron is always spread all over the space. Moreover, according to  \cite{H85}  there holds the limited spatial decay
$\int_{\{|x|>r\}} |\Psi(x)|^2\d^3x$ $\notin \mathcal{O}(\operatorname{e}^{-Kr})$, $r\to \infty$  for $K >2m$.
\\
\hspace*{6mm}
 So $\int_\Delta |\Psi(x)|^2\d^3x<1$ for every bounded region $\Delta$ or even every closed $\Delta\ne \R^3$. Nevertheless the Dirac electron is localizable within every however small ball, not strictly but as accurately as desired. Indeed,  for every point $b\in\R^3$ there is  a sequence $(\Psi^{(n)})$  of normalized electron wavefunctions  localized at $b$, which means $\int_B|\Psi^{(n)}(x)|^2\d^3x\to 1$, $n\to\infty$ for every ball $B$ around $b$
\cite{BFM}, \cite[sec.\,G,H]{CL15}. Moreover, one has the causal behavior that, at every time $t$, $\int_B|\Psi_{t}^{(n)}(x)|^2\d^3x\to 1$, $n\to\infty$  for every ball $B$ around $b$ with radius $> |t-t_0|$  \cite[(16) Theorem]{C18}. 
\\
\hspace*{6mm}
Localization by means of point-localized sequences of states is closely related to positive operator localization  (used by many authors as initially Neumann, Castrigiano, Kraus, and others), which is a generalization of WL  based on an unconventional notion of observable called {\it effect} and {\it generalized observable} by Ludwig, or {\it fuzzy observable} by Ali and Emch, or {\it unsharp observable} by Busch et al. For details see  \cite[sec.\,F,G]{CL15} and \cite[sec.\,6,8,15]{C18}. In case of the Dirac electron the unsharp localization is just the trace of the canonical localization of the Dirac system on the subspace of positive energy.
\\
\hspace*{6mm}
There is an important implication of point-localized sequences of states, which puts the bounded localized states into perspective. Recall that by Hegerfeldt's theorem  an   admixture of unbounded negative energy is needed to localize the Dirac system in a bounded region. However, if there is $(\Psi^{(n)})$  localized at $b$, then the amount  of negative energy needed to localize the system in any  ball $B$ around $b$  is arbitrarily small.  More precisely, there is a sequence of normalized Dirac wavefunctions $\psi^{(m)}$ such that $1_B \psi^{(m)}=\psi^{(m)}$ up to finitely many $m$  for every ball $B$ around $b$ and such that $(I-P)\psi^{(m)}\to 0$, $m\to \infty$, with $1_B$ the indicator function of $B$ and $P$  the orthogonal projection on the subspace of positive energy. 
\\
\hspace*{6mm}
Quantitative results on the above mentioned admixture of negative energy are important. Quite generally, given any WL $E$ and an orthogonal projection $P$ with non-vanishing dilational limit, then  the above result holds  with $E(B) \psi^{(m)}$ in place of $1_B \psi^{(m)}$  \cite[Theorem 7, Lemma 7]{CL15}. Recently \cite{FP20}, a detailed investigation is addressed to the proof of quantitative versions of Hegerfeldt's theorem including results on the energy spectrum of bounded localized wavefunctions.\\
\hspace*{6mm}
Plainly, the presence of negative energy means that the antiparticle positron comes into play. In our view a Dirac state $\psi$ is a superposition of an electron and a positron state. However, it is only virtual as suppressed by the relevant superselection rule. Only when a measurement is performed the state after is a real mixed state of electron and positron states.  After a position measurement regarding a bounded region $\Delta$ of the Dirac  system in an electron state,  the resulting electron and positron states obviously are not localized in $\Delta$.  Hence as derived in  \cite[sec.\,J]{CL15} the attempt to localize an electron leads to the 
 pair creation of non-localized real particles. This mechanism is often considered to be the true obstruction of particle localization in relativistic quantum mechanics. For a brief reflection in a  field-theoretic context see \cite{H98}.
 \\
\hspace*{6mm}
So the  wavefunctions with non-semi-bounded energy spectrum, above all the bounded localized ones,  play an essential role  in causal particle localization. We will study the Dirac time evolution in order to gain insight into their causal behavior. Let us  describe the outcomes. The free Dirac operator in position representation is
\begin{equation}\label{DOPR}
H=\sum_{k=1}^3\alpha_k\,\frac{1}{\i} \partial_k+ \beta m  
\end{equation}
acting in $L^2(\R^3,\C^4)$, where the units are such that $c=1$, $\hbar=1$.  If $\psi$
 is the wavefunction at time $0$, then the time-translated wavefunction  by $t$   is 
$\psi_t=\e^{\i tH}\psi$. The Dirac time evolution is  causal. This means, if $\psi$ is localized in the region (measurable subset)  $\Delta\subset \R^3$, i.e., $1_\Delta\psi=\psi$ a.e., then $\psi_t$  is localized in  the region of influence, 
which is the set of points  reached from $\Delta$  within time  $|t|$ at the speed of light. Usually, as in \cite[sec.\,1.5]{T92}, this is inferred from the fact that the propagator depends only on $\operatorname{sgn}(t)$ and $t^2-x^2$ and vanishes if $t^2-x^2<0$  \cite[(1.86)]{T92}. Here,  we like to cite  \cite[Theorem 10(b)]{CL15}, which infers causal time evolution  from the fundamental  fact that the entire matrix-valued function $z\mapsto \operatorname{e}^{\i th(z)}$ on $\C^3$ with
\begin{equation}\label{CEE}
h(z):=\sum_{k=1}^3\alpha_k z_k + \beta m
\end{equation}
 is exponentially bounded.\\
\hspace*{6mm}
Causality together with homogeneity of time gives a first idea of how Dirac wavefunctions propagate in space. The spreading to infinity all over the space is limited by the velocity of light. However causality implies also the non-superluminal shrinking of the carrier of  wavefunctions as the following simple consideration tells.  Let $R>0$ and let $\psi$ be a wavefunction localized in the  ball $B_R:=\{x\in\R^3:|x|\le R\}$. 
Let $t_0:=R$. Then, due to causality,  $\psi_{t_0}$ is localized in $B_{2R}$, but this does not exclude that  actually  $\psi_{t_0}$ is localized in a smaller ball $B_r$. Indeed, here  every $r>0$  occurs: Choose $\rho\in ]0,R[$, and let the wavefunction $\chi\ne 0$ be localized in $B_\rho$,  
then $\psi:=\chi_{\rho-R}$ is localized in $B_R$, and $\psi_{t_0}=\chi_{t_0+\rho-R}=\chi_\rho$
is localized in $B_{2\rho}$. \\
\hspace*{6mm}
Indeed, the phase of shrinking of the carrier is not accidental. Exploiting  further (\ref{CEE}) in sec.\,\ref{PBWF} it turns out that in the past the carrier of every  bounded localized wavefunction shrinks isotropicly at light speed. Subsequently it expands  to infinity in the same manner.  Obviously this kind of movement does not single out some direction of time. On the contrary,  the reversal of motion is required by time reversal symmetry. The phase of rebound is particularly interesting.  Limited temporally and spatially  in the order of the diameter of the carrier at its minimal extension it constitutes a complicated movement. For every direction in space  there is a definite time, at which like a bounce  the change  from shrinking to expanding happens abruptly. In this respect this feature reminds of the  phenomenon of the zitterbewegung.  However the  motion is not superposed by the zitterbewegung. 
This intriguing temporal behavior of the carrier, shown in (\ref{SETIIED}),  is not known up to now.  Like the zitterbewegung, the rebound motion is a relativistic quantum phenomenon. 
\\
\hspace*{6mm}
A further aspect of time evolution concerns the long-term behavior of the position probability density $|\psi_t(x)|^2$ across the carrier of a wavefunction $\psi$. In sec.\,\ref{LTBPL} it is shown  that   in the past as in the future the probability of localization concentrates up to $1$ in the spherical  shell $B_{|t|}\setminus B_{r}$ for every radius $r>0$. In conclusion  the so-called asymptotic causality is briefly discussed.\\
\hspace*{6mm}
In the sections \ref{PBWF} and \ref{LTBPL} the results are presented. Their proofs are postponed to sec.\,\ref{P}.
For $x,y\in\R^3$ put $xy:=\sum_{k=1}^3x_ky_k$. $\mathcal{F}$ denotes the unitary Fourier transformation on $L^2(\R^d,\C^n)$. For open $U\subset \R^d$, $\mathcal{C}^\infty_c(U,\C^n)$ is the space of  all infinitely differentiable functions on $\R^d$ in $\C^n$ with compact support in $U$.

\section{Motion of the border of the wavefunction}\label{PBWF}

\begin{Def}\label{BD}  Let  $e\in\R^3$  be a unit vector   and  $\alpha\in[-\infty,\infty]$. They determine the half-space 
$\{x\in\R^3: x\,e\le\alpha\}$ (which equals  $\emptyset$ or $\R^3$ if $\alpha=-\infty$ or $\alpha=\infty$). For every $\psi\in L^2(\R^3,\C^4)\setminus\{0\}$ let $e(\psi)\in[-\infty,\infty]$ denote the maximal $\alpha$ satisfying $1_{\{x\in\R^3: x\,e\le\alpha\}}\psi=0$ a.e.  Put $\overline{e}:=-e$. 
\end{Def}

 The meaning of $e(\psi)$ is best elucidated by

\begin{Lem}\label{EEP} $\psi\ne 0$   is localized in
$\{x\in\R^3: e(\psi)\le xe \le -\overline{e}(\psi)\}$
with $]e(\psi), -\overline{e}(\psi)[$  the smallest interval with this property. In particular, $ -\overline{e}(\psi)-e(\psi)>0$ is the width of the carrier of $\psi$ in direction $e$.
\end{Lem}

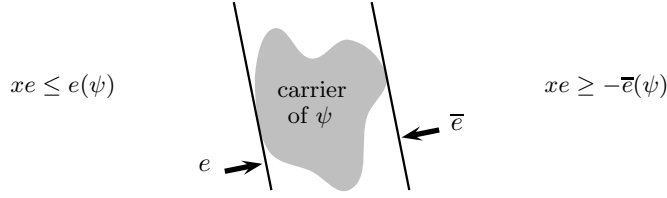
\begin{center}
\begin{pspicture}(3.5,2.5)
\pscurve*[linecolor=lightgray](0.25,1.5)(0.5,2.15)(1,1.75)(1.5,2)(2,1.5)(1.75,1)(1.5,0)(1,0.25)(0.45,0.5)(0.25,1.5)

\put(0.69,0.89){\small{of $\psi$}}
\put(0.55,1.24){\small{carrier}}
\put(-3,1.3){\small{$xe\le e(\psi)$}}
\put(4.1,1.3){\small{$xe\ge -\overline{e}(\psi)$}}
\psline[linewidth=2pt]{->}(-0.15,0.25)(0.35,0.35)
\put(-0.5,0.25){$e$}

\psline[linewidth=2pt]{<-}(2.2,0.7)(2.7,0.8)
\put(2.85,0.8){$\overline{e}$}

\linethickness{0.3mm}
\put(0.47,0){\line(-1,5){0.5}}
\put(2.3,0){\line(-1,5){0.5}}
\end{pspicture}
\end{center} 

\vspace{5mm}
In the following we are interested in the temporal behavior of $e(\psi)$, i.e., in the functions $\R\to \R$, $t\mapsto e(\psi_t)$. 

\begin{The}\label{GGETIIED} Let $\psi\ne 0$ be a Dirac wavefunction localized in a bounded region. Then 
\begin{equation*} e(\psi_t)\le -2\,\overline{e}(\psi)-e(\psi) -|t| 
\end{equation*}
holds for all directions $e$ and all times $t$. If  $\psi\in \operatorname{dom}(H)$ or if more generally
$h\,\mathcal{F}\psi$ is bounded
on $\R^3$ then the inequality holds even with $<$  in place of $\le$.
\end{The}

For the proof of (\ref{SETIIED}), 
Theorem (\ref{GGETIIED}) turns out to be  decisive as it shows that $t\mapsto e(\psi_t)$ is bounded above.
Also, together with (\ref{SETIIED})  it implies the important estimations in (\ref{CSETIIED}). Afterwards the  bound $-2\,\overline{e}(\psi)-e(\psi) $ in (\ref{GGETIIED}) can be improved to $-2\,\overline{e}(\psi)-e(\psi)- |t_{\overline{e}}| $
by   (\ref{CSETIIED})(a),(b) and (\ref{SETIIED}).

\begin{The}\label{SETIIED} Let $\psi\ne0$ be a Dirac wave function localized in a bounded region. Then there exists a unique time $t_e=t_e(\psi)\in\R$ such that 
\begin{equation*}  e(\psi_t)= e(\psi)+|t_e|-|t-t_e|
\end{equation*}
for all times $t\in\R$ and directions $e$. In particular $e(\psi_{t_e})=e(\psi)+|t_e|$ is the maximum of $e(\psi_t)$.
\end{The}\\

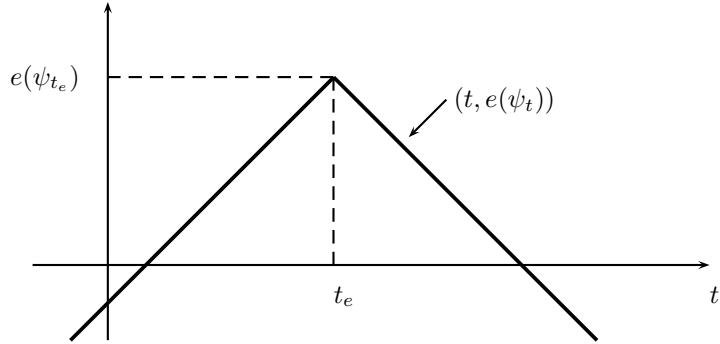
\begin{pspicture}(-6,-1)(3,3.5)
\psline{->}(-2,-1)(-2,3.5)
\psline{->}(-3,0)(6,0)
\psline{->}(2.5,2.2)(2,1.7)
\put(2.6,2.1){$(t,e(\psi_t))$}
\linethickness{0,5mm}
\put(1,2.5){\line(-1,-1){3.5}}
\put(1,2.5){\line(1,-1){3.5}}
\psline[linestyle=dashed](1,2.5)(1,0)
\psline[linestyle=dashed](1,2.5)(-2,2.5)
\put(1,-0.5){$t_e$}
\put(6,-0,5){$t$}
\put(-3.3,2.4){$e(\psi_{t_e})$}
\end{pspicture}

\begin{Cor}\label{CSETIIED} Let $\psi\ne 0$ be a Dirac wavefunction localized in a bounded region. Then
\begin{itemize}
\item[\emph{(a)}] $e(\psi_t)\le -\overline{e}(\psi)-|t-t_e|$
\item[\emph{(b)}] $|t_e|+|t_{\overline{e}}|\le -\overline{e}(\psi)-e(\psi)$
\item[\emph{(c)}] $2|t_e|<-\overline{e}(\psi)-e(\psi)$ if $t_e=t_{\overline{e}}$
\item[\emph{(d)}]   $t_e(\psi_t)=t_e(\psi)-t$ for $t\in\R$
\item[\emph{(e)}]  If $h\,\mathcal{F}\psi$ is bounded
on $\R^3$ then the inequalities in \emph{(a), (b)} hold even with $<$  in place of $\le$.
\end{itemize}
\end{Cor}
\begin{Cor}\label{LTSIE} Let $\psi\ne 0$ be a Dirac wavefunction localized in $B_R$ for some $R>0$.  Then 
$$ e(\psi_t)=e(\psi_{2R})+2R-t \quad \forall\; t\ge 2R, \quad e(\psi_t)=e(\psi_{-2R})+2R+t \quad \forall\; t\le -2R$$
\end{Cor}\\
 So  the carrier of a bounded localized Dirac wavefunction $\psi$ performs an intriguing motion.  As long as $t<t_e$, one has
$e(\psi_t)=e(\psi_{t_e})-t_e +t$ by (\ref{SETIIED}), which means the retreat at the speed of light of the carrier of $\psi_t$ in  direction $e$. Only after time $t_e$ the carrier advances in direction $-e$ at the speed of light as $e(\psi_t)=e(\psi_{t_e})+t_e -t$. Only then the wavefunction expands in the direction $-e$ as expected. The abrupt change at the time $t_e$ of the direction of  the motion with light velocity  to the opposite direction is like a bounce.
It reminds of the  phenomenon of the zitterbewegung. But this behavior  is easy to understand. Let $\psi':=\psi_{t_e}$. Then by (\ref{CSETIIED})(d), which is due to homogeneity of time, i.e., the translational symmetry of time evolution, $\psi'$ satisfies $e(\psi'_t)=e(\psi')-|t|$ according to (\ref{SETIIED}). So, as maximal permissible by causality, $\psi'$ expands in the future as well in the past  in direction $-e$ at the speed of light. In particular the result in (\ref{SETIIED}) does not single out some direction of time. On the contrary,  the reversal of motion is required by time reversal symmetry.
Nevertheless for a short period  the picture is complicated as  the time of change $t_e$ depends in general on the direction $e$ (see (\ref{NISE})). 
Therefore the carrier of the wavefunction performs the change from shrinking to expanding not isotropicly.  According to  (\ref{CSETIIED})(a), in every direction $e$ the retreat equals at most the width of the carrier. Moreover, after and  before the time corresponding to the diameter of the carrier, a simultaneous  isotropic expansion   of the wavefunction  with light velocity takes place in the future and in the past, respectively (see (\ref{LTSIE})). Thus the motions of the parts of the border result in a rebound of the wavefunction. The phase of rebounding is limited in time and space  in the order of the diameter of the carrier at its minimal extension.\\

This section is concluded by two existence  proofs for  bounded localized  wavefunctions $\psi\ne0$
regarding the data $e(\psi)$, $t_e$, $\overline{e}(\psi)$, $t_{\overline{e}}$. Fix a direction $e$.

\begin{Lem}\label{NISE} 
  Let $t_1,t_2 \in \R$. Then there is a bounded localized  wavefunction $\psi\ne0$ with $t_e=t_1$ and $t_{\overline{e}}=t_2$.
 \end{Lem}\\
This means that the shrinking-expanding point may take place in direction $e$ and  opposite direction $-e$ at different times causing an anisotropic movement of the wavefunction as described above.

\begin{Lem}\label{DSABTP}
For every $a,b\in\R$ with $a<b$ and $ |\tau| <\frac{1}{2}(b-a)$ 
 there is a bounded localized  wavefunction $\psi\ne0$  such that $a\le e(\psi)<-\overline{e}(\psi)\le b$  and $t_e=t_{\overline{e}}=\tau$.
\end{Lem}\\
So the estimation given  in (\ref{CSETIIED})(c) cannot be improved.\\ 
\hspace*{6mm}
Wavefunctions satisfying  $e(\psi)\ge0$, $2\,|t_{\overline{e}}|\ge - \overline{e}(\psi)$ are particularly interesting.   Let us remark that using properties of these wavefunctions,   one may show the phenomenon of Lorentz contraction, i.e.,  for every Dirac wavefunction $\psi$ and every   $\delta>0$
$$\norm{1_{\{x\in\R^3: -\delta\le xe \le \delta\}}\,\psi^{\rho e}}\, \to 1, \;\; |\rho|\to\infty$$
holds, where $\psi^{\rho e}$ denotes the wavefunction $\psi$ boosted in direction $e$ with rapidity $\rho\in \R$.

\section{Long-term behavior of the probability of localization}\label{LTBPL} Let $\psi$ be  a normalized Dirac wave function.    
Obviously the probability of localization within the carrier of the  wave function evolving in time  is constantly $1$. Insomuch the foregoing results on the movement of the border of the wave function  yield also an information about the time dependence of the probability of localization. However the probability stays not equally distributed across the carrier. It turns out that for every Dirac wavefunction (not necessarily bounded localized) in the long term the probability of localization concentrates up to $1$ in the spherical  shell $B_{|t|}\setminus B_{r}$ for every radius $r>0$. More precisely one has  the results in (\ref{PRBG}), (\ref{VPOBT}).

\begin{The}\label{PRBG} Let $\psi$ be a Dirac wavefunction. Let $\varepsilon> 0$. Then there are $v\in\,]0,1[$ and $\tau>0$ such that $\norm{1_{B_{v |t|}}\psi_t}\le \varepsilon$ for all $|t|\ge \tau$. In particular
$$1_{B_r}\psi_t\to 0, \quad |t|\to \infty$$ holds for every radius $r>0$.
\end{The}\\

\begin{The}\label{VPOBT}  Let $\psi$ be a Dirac wavefunction. Then 
$$ 1_{\R^3\setminus B_{|t|}}\psi_t\to 0, \;\;|t|\to \infty$$
If $\mathcal{F}\psi\in \mathcal{C}^\infty_c(\R^3,\C^4)$ holds, then for every $N>0$ there is a finite constant $C_N$ such that $\norm{1_{\R^3\setminus B_{|t|}}\psi_t}\le C_N(1+|t|)^{-N}$ for all $t\in\R$.
\end{The}

\begin{AC} \label{AC} Actually both  results (\ref{PRBG}), (\ref{VPOBT}) are valid  also for every massive system and antisystem $[m,j,\eta]$ ($m>0$, spin $j\in\mathbb{Z}/2$, $\eta=+,-$) if endowed with the Newton-Wigner localization $E^{\textsc{nw}}$ \cite{W62}, although with respect to $E^{\textsc{nw}}$ time evolution is not causal. (The  result corresponding to (\ref{VPOBT}) is known for a long time \cite{WM63} and studied in detail in \cite{R81}.)
So one has $E^{\textsc{nw}}(B_r)\psi_t\to 0, \;\;|t|\to \infty$ for every $r>0$ and   the {\it asymptotic causality}
\begin{equation}\label{ASC}
 E^{\textsc{nw}}(\R^3\setminus B_{|t|})\psi_t\to 0, \;\;|t|\to \infty
 \end{equation}  
Indeed, these results hold true since the evolution of a state  $\psi$  in Newton-Wigner position representation is 
$ \psi_t(x)=(2\pi)^{-3/2}\int \operatorname{e}^{\operatorname{i}(px+t\eta\epsilon(p))}\big(\mathcal{F}\psi\big)(p)\,\operatorname{d}^3p$,
 which  for every spinor component equals (\ref{ENSPM}). \\
\hspace*{6mm}
The asymptotic causality of Newton-Wigner localization is  shown in   \cite[Proposition]{R81}. 
 In \cite{R81} it is also pointed out that (\ref{ASC}) is false for the massless system $[0,0,\eta]$. But the failure of (\ref{ASC}) must not mean at all an  acausal behavior. Indeed,  although  radially symmetric Weyl wavefunctions  satisfy $\lim_{|t|\to\infty}\norm{1_{\R^3\setminus B_{|t|}}\psi_t}\ge 1/2$,   the  Weyl systems are causal 
 \cite[(99)(b), (95)]{C18}.
\end{AC}

\section{Proofs}\label{P}

\textbf{Proof of\, (\ref{EEP})\,Lemma.} Note that $\{x\in\R^3: x\operatorname{e}=\gamma\}$ is a Lebesgue null set.  
By definition $1_{\{x\in\R^3: x\,e \le\alpha\}}\psi=0$, $1_{\{x\in\R^3: x\,\overline{e}\le \beta\}}\psi=0$ exactly for all $\alpha\le e(\psi)$ and $\beta \le  \overline{e}(\psi)$. From this it follows $1_{
\{x\in\R^3: e(\psi)\le xe \le -\overline{e}(\psi)\}}\psi=\psi$, whence the assertion.\qed\\

The proof of (\ref{GGETIIED}) needs some preparation.
Referring to (\ref{BD}) define  $e(\eta)$ for $\eta\in L^2(\R^3,\C)$ quite analogously. Obviously $e(\psi)=\min_le(\psi_l)$. --- Recall the  \textbf{support function}  $H_C$ for a convex set $C\subset \R^d$ given by
 \begin{equation}\label{SFCS} 
  H_C(\lambda)=\sup\{x\lambda:x\in C\},\, \lambda\in\R^d
 \end{equation}

Let $C(\psi)$ denote the smallest convex set outside which $\psi$ vanishes almost everywhere. Clearly, $\{x\in\R^3: x\,e\le e(\psi)\}\cap C(\psi)=\emptyset$ and $\{x\in\R^3: x\,e\le \alpha\}\cap C(\psi)$ is not a null set if $\alpha>e(\psi)$. Hence $e(\psi)=\inf\{xe:x\in C(\psi)\}$.  These considerations are applicable as well  to every component $\psi_l$ of $\psi$. 
Therefore 
\begin{equation}\label{REH}
e(\psi)=-H_{C(\psi)}(-e), \quad e(\psi_l)=-H_{C(\psi_l)}(-e)
\end{equation}

The \textbf{P-indicator}  (i.e., the P\'olya-Plancherel indicator) $h_f$ of an entire function $f$ on $\C^d$ is 
\begin{equation}
 h_f(\lambda)=\sup\{h_f(\lambda,x):x\in\R^d\},\, \lambda\in\R^d \textrm{\, with\, } h_f(\lambda,x)=\varlimsup_{r\to \infty}\frac{1}{r}\ln |f(x+\operatorname{i}\lambda \,r)| 
 \end{equation}
An entire  matrix-valued function $f$ on $\C^d$ is called \textbf{exponentially bounded} or of exponential type with exponent $\delta\ge 0$ if there is a finite constant $C_\delta$ such that $\norm{f(z)}\le C_\delta \operatorname{e}^{\delta|z|}$ with $|z|^2=\sum_{j=1}^d|z_j|^2$ for  $z\in\C^d$. The \textbf{type} $\tau$ of $f$ is the infimum of all its exponents. \\
 \hspace*{6mm}
The main mathematical tool for the proof of (\ref{GGETIIED}) is the Theorem of   Plancherel and P\'olya and related results   
(see \cite{R86},  \cite{R74}), which for convenience we put together here 

\begin{TPP}\label{TPP}   
  {\it A function $f:\C^d\to\C$ is entire and exponentially bounded  
  with $f|_{\R^d}\in L^2$ if and only if there is $g\in L^2(\R^d)$ vanishing outside a bounded set with
 \begin{equation*}  
 f(z)=(2\pi)^{-d/2}\int_{\R^d}\operatorname{e}^{-iqz}g(q)\operatorname{d}q 
\end{equation*}
 i.e., $f$ is the Fourier-Laplace transform of $g$. 
  Then  $$h_f=H_{C(g)}$$
   where $C(g)$ is the smallest convex set outside of which $g$ vanishes almost everywhere. \\
 \hspace*{6mm} 
  Moreover,  $f|_{\R^d}$  is bounded  by $(2\pi)^{-d/2}\int_{\R^d}|g(q)|\operatorname{d}q$ and, by the Riemann-Lebesgue lemma, 
  it  vanishes at infinity. Also,
  for each $\lambda$ one has $h_f(\lambda,x)=h_f(\lambda)$ for almost all $x$, and $h_{kf}=h_k+h_f$ holds for  any exponentially bounded entire function $k$.} 
\end{TPP} \\ 
Let $\psi$ be a wavefunction localized in $B_R$. The Fourier-Laplace transform  of $\psi_t$ localized in $B_{R+|t|}$ is an entire function $\varphi_t$,  which 
 is exponentially bounded with exponent $R+|t|$, i.e., $|\varphi_t(z)|\le C\operatorname{e}^{(R+|t|)|z|}$, $z\in\C^3$.  Recall (\ref{CEE}). Due to $h(z)^2=(z^2_1+z^2_2+z^2_3+m^2)I_4$, the time evolution yields 
 \begin{equation}\label{TED}
\varphi_t(z)=\operatorname{e}^{\operatorname{i}th(z)}\varphi(z)=\cos\big(t\epsilon(z)\big)\,\varphi(z)+\operatorname{i}t\operatorname{sinc}\big(t\epsilon(z)\big)\,h(z)\varphi(z)
\end{equation}
for every $z\in\C^3$.  Here $\epsilon$ satisfies $\epsilon(z)^2=z_1^2+z_2^2+z_3^2+m^2$, and $\operatorname{sinc}(w)=\sin(w)/w$ for $w\ne 0$, $\operatorname{sinc}(0)=1$, is the  sine cardinal.  From (\ref{TED}) one obtains 
\begin{equation}\label{STED}
\varphi_t+\varphi_{-t}=2\cos(t\epsilon)\,\varphi, \quad \varphi_t-\varphi_{-t}=2\operatorname{i}t\operatorname{sinc}(t\epsilon)\,h\varphi
\end{equation}
and
$(\varphi_t)_k\varphi_l=\cos(t\epsilon)\varphi_k\varphi_l+\operatorname{i}t\operatorname{sinc}(t\epsilon)(h\varphi)_k\varphi_l$ and hence
$\phi_{kl}=\operatorname{i}t\operatorname{sinc}(t\epsilon)\chi_{kl}$
  for $\phi_{kl}:=(\varphi_t)_k\varphi_l-(\varphi_t)_l\varphi_k$ and  $\chi_{kl}:=(h\varphi)_k \varphi_l-(h\varphi)_l \varphi_k$, where $k,l=1,\dots,4$ enumerate the spinor components. \\
 \hspace*{6mm}
 There are  $k,l$  such that $\chi_{kl}\ne 0$. (Indeed, assume $\chi_{kl}=0$ for all $k,l$. Then $\varphi_k\, h\varphi=(h\varphi)_k\, \varphi$ and hence $\epsilon^2\,\varphi_k\,\varphi=(h\varphi)_k\, h\varphi$. Then $(h\varphi)_k^2-\epsilon^2\,\varphi_k^2=0$. Fix $k,z_2,z_3$ such that $f(\zeta):=\varphi_k(\zeta,z_2,z_3)$ is not the null function. Set $g(\zeta):=(h\varphi)_k(\zeta,z_2,z_3)$. Choose the square root $\mu(\zeta)$ of $\zeta^2+z_2^2+z_3^2$ such that $g=\mu f$. This, however, is impossible as $g/f$ is meromorphic whereas $\mu$ is not.)
Thus $\chi:=\chi_{kl}$, $\phi:=\phi_{kl}$ are non-zero entire exponentially bounded functions with exponents $2R$ and $2R+|t|$, respectively, satisfying
\begin{equation}\label{DRDTE}
\phi=\operatorname{i}t\operatorname{sinc}(t\epsilon)\,\chi
 \end{equation} 
 We are going to exploit the relations (\ref{STED}) and (\ref{DRDTE}). The following elementary but not trivial estimations are used  to compute the P-indicator for $\cos(t\epsilon)$ and $\operatorname{sinc}(t\epsilon)$ in (\ref{PICESCE}).
 
\begin{Lem}\label{EFSINC} Let $\mu, t,u,v$ be real, $\mu\ge 0$. Then there are  finite constants $A_t>0,B_t,C_t$ independent of $u,v$ such that 
\begin{itemize}
\item[\emph{(a)}] $A_t \operatorname{e}^{|t v|}\le\big| \cos\big(t\sqrt{\mu^2+(u+\operatorname{i}v)^2}\,\big)\big|\le B_t\operatorname{e}^{|t v|}$
\item[\emph{(b)}] $A_t  |u+\operatorname{i}v|^{-1}\operatorname{e}^{|t v|}\le\big| \operatorname{sinc}\big(t\sqrt{\mu^2+(u+\operatorname{i}v)^2}\,\big)\big|\le B_t  |u+\operatorname{i}v|^{-1} \operatorname{e}^{|t v|}$
\end{itemize}
for all $u$ and  $|v|>  C_t$. 
\end{Lem}

{\it Proof.}
 First we show
\begin{equation*} 
z\in \C, |z|\le \frac{1}{2}\; \Rightarrow \;\sqrt{1+z}=1+\zeta \textrm{\; with\; }  |\zeta|\le \frac{3}{4}|z|\tag{$\star$}
\end{equation*}
Indeed, let $f:[0,1]\to \C$, $f(r):=(1+r z)^{1/2}$. As $f'(r)=\frac{1}{2}(1+r z)^{-1/2}z$ one has $f(1)=1+\zeta$
with  $\zeta:=\int_0^1f'(r)\operatorname{d}r$  and $|\zeta|\le 1\cdot \frac{1}{2}(1-\frac{1}{2})^{-1/2}|z|$, whence ($\star$).\\
\hspace*{6mm}
Now  assume at once $t\ne 0$ and let in the following $|v|> \sqrt{2}\mu$. Put $s(u,v):=\sqrt{\mu^2+(u+\operatorname{i}v)^2}$. More precisely, 
$s(u,v):= (u+\operatorname{i}v)\sqrt{1+z}$ for $z:=\mu^2(u+\operatorname{i}v)^{-2}$ with $|z|=\mu^2(u^2+v^2)^{-1}\le\mu^2v^{-2}\le\frac{1}{2}$. \\
\hspace*{6mm}
By ($\star$), 
$s(u,v)=
(u+\operatorname{i}v)(1+\zeta)$ with $|u+\operatorname{i}v|\,|\zeta|\le \frac{3}{4}|u+\operatorname{i}v|\,|z|\le
\mu^2|v|^{-1}\le\mu$. This implies $|\operatorname{e}^{\pm\operatorname{i}ts(u,v)}|=
|\operatorname{e}^{\pm\operatorname{i}t(u+\operatorname{i}v)(1+\zeta)}|=\operatorname{e}^{\mp tv}|\operatorname{e}^{\pm\operatorname{i}t(u+\operatorname{i}v)\zeta}|$. One concludes
\begin{equation*}\operatorname{e}^{-|t|\mu}\operatorname{e}^{-tv}\le |\operatorname{e}^{\operatorname{i}ts(u,v)}|\le \operatorname{e}^{|t|\mu}\operatorname{e}^{-tv}, \quad \operatorname{e}^{-|t|\mu}\operatorname{e}^{tv}\le |\operatorname{e}^{-\operatorname{i}ts(u,v)}|\le \operatorname{e}^{|t|\mu}\operatorname{e}^{tv}\tag{$\star\star$}
\end{equation*}
for  all $v\in\R$ with $|v|> \sqrt{2}\mu$.  

\hspace*{6mm}
(a) $|\cos(w)|=\frac{1}{2}|\operatorname{e}^{\operatorname{i}w}+\operatorname{e}^{-\operatorname{i}w}|\le \frac{1}{2}(|\operatorname{e}^{\operatorname{i}w}|+|\operatorname{e}^{-\operatorname{i}w}|)$. For $w=ts(u,v)$ this yields by ($\star\star$) $|\cos(ts(u,v))|\le\frac{1}{2}( \operatorname{e}^{|t|\mu}\operatorname{e}^{-tv}+ \operatorname{e}^{|t|\mu}\operatorname{e}^{tv})\le \operatorname{e}^{|t|\mu}\operatorname{e}^{|tv|}$. Hence the right part of the  inequality of (a) holds for $B_t:=\operatorname{e}^{|t|\mu}$ and $C'_t=\sqrt{2}\mu$.\\
\hspace*{6mm}
For the left part of the inequality use $|\cos(w)|=\frac{1}{2}|\operatorname{e}^{\operatorname{i}w}+\operatorname{e}^{-\operatorname{i}w}|\ge \frac{1}{2}\big||\operatorname{e}^{\operatorname{i}w}|-|\operatorname{e}^{-\operatorname{i}w}|\big|$. Then for $w=ts(u,v)$ one gets by ($\star\star$) $|\cos(ts(u,v))|\ge\frac{1}{2}( \operatorname{e}^{-|t|\mu}\operatorname{e}^{|tv|}- \operatorname{e}^{|t|\mu}\operatorname{e}^{-|tv|})=\sinh\big(|t|(|v|-\mu)\big)$. Check $\sinh(x)\ge\frac{1}{4}\operatorname{e}^x$ for $x\ge \frac{\ln(2)}{2}$. Thus we conclude that the left part of the inequality holds for $A_t:=\frac{1}{4}\operatorname{e}^{-|t|\mu}$ and $C''_t:=\sqrt{2}\mu+\frac{\ln(2)}{2}\frac{1}{|t|}$.\\
\hspace*{6mm}
(b) Check first $|t\,s(u,v)|\ge \frac{|t|}{\sqrt{2}}|u+\operatorname{i}v|$, using $|\sqrt{1+z}|\ge \frac{1}{\sqrt{2}}$ for $|z|\le\frac{1}{2}$. Furthermore, $ |\sin(w)|=\frac{1}{2}|\operatorname{e}^{\operatorname{i}w}-\operatorname{e}^{-\operatorname{i}w}|\le \frac{1}{2}(|\operatorname{e}^{\operatorname{i}w}|+|\operatorname{e}^{-\operatorname{i}w}|)$. Hence,  as for (a),  the right part of the inequality holds for $B_t:=\frac{\sqrt{2}}{|t|}\operatorname{e}^{|t|\mu}$ and $C'_t=\sqrt{2}\mu$.\\
\hspace*{6mm}
Regarding the left part of the inequality of (b), we estimate $|ts(u,v)|^{-1}\ge (\frac{2}{3})^{1/2} |t(u+\operatorname{i}v)|^{-1}$, as $|s(u,v)|=|u+\operatorname{i}v|\,|\sqrt{1+z}\,|\le(\frac{3}{2})^{1/2} |u+\operatorname{i}v|$.  Furthermore, one has $|\sin(w)|=\frac{1}{2}|\operatorname{e}^{\operatorname{i}w}-\operatorname{e}^{-\operatorname{i}w}|\ge \frac{1}{2}\big||\operatorname{e}^{\operatorname{i}w}|-|\operatorname{e}^{-\operatorname{i}w}|\big|$.  Hence, proceeding as in (a), it follows that the left part of the inequality holds for $A_t:=(\frac{1}{24})^{1/2}\frac{1}{|t|}\operatorname{e}^{-|t|\mu}$ and $C''_t:=\sqrt{2}\mu+\frac{\ln(2)}{2}\frac{1}{|t|}$.
\qed\\

\begin{Lem}\label{PICESCE} For $t\in\R$ the functions $z\mapsto \cos\big(t\epsilon(z)\big)$ and $z\mapsto \operatorname{sinc}\big(t\epsilon(z)\big)$ are bounded on $\R^3$ and entire on $\C^3$ with exponent $|t|$, which is minimal. Moreover, $h_{\cos(t\epsilon)}(\lambda)=h_{\operatorname{sinc}(t\epsilon)}(\lambda)=|t|\,|\lambda|$ holds for $\lambda\in\R^3$. More precisely one has $|t|\,|\lambda|=\lim_{r\to\infty}\frac{1}{r}\ln|f(p+\operatorname{i}\lambda r)|$, $p\in\R^3$ for  $f\in\{\cos(t\epsilon),\operatorname{sinc}(t\epsilon)\}$. 
 \end{Lem}
 
{\it Proof.}
We show the assertion for  $\operatorname{sinc}(t\epsilon)$. Regarding $\cos(t\epsilon)$ the proof is analogous. Assume at once $t\ne 0$.\\
\hspace*{6mm}
Obviously, $\operatorname{sinc}(t\epsilon)$ is bounded on $\R^3$ and entire on $\C^3$. Also, there is an entire function $s$ satisfying $s(z^2)=\operatorname{sinc}\big(t\epsilon(z)\big)$ with $z^2=z_1^2+z_2^2+z_3^2$ for all $z\in\C^3$. ---
Now $|\epsilon(z)|^2=|z^2+m^2|\le |z^2|+m^2 = |z|^2+m^2\le(|z|+m)^2$, whence $|\epsilon(z)|\le |z|+m$. Therefore, $|\sin\big(t\epsilon(z)\big)|=
\frac{1}{2}|\operatorname{e}^{\operatorname{i}t\epsilon(z)}-\operatorname{e}^{-\operatorname{i}t\epsilon(z)}|\le \operatorname{e}^{|t|\,|\epsilon(z)|}\le \operatorname{e}^{|t|m} \operatorname{e}^{|t|\,|z|}$ for all $z$.\\
\hspace*{6mm}
If $|z^2|\le 2m^2+1$ then  $|\operatorname{sinc}\big(t\epsilon(z)\big)|=|s(z^2)|\le C$ for some finite constant $C$. For $|z^2|> 2m^2+1$ one has $|\epsilon(z)|^2=|z^2+m^2|\ge |z^2|-m^2>m^2+1$, whence $|\epsilon(z)|^{-1}<1$.
Hence $|\operatorname{sinc}\big(t\epsilon(z)\big)|\le C'\operatorname{e}^{|t|\,|z|}$ for all $z$, where $C':=C+\frac{\operatorname{e}^{|t|m}}{|t|}$. So $|t|$ is an exponent for $\operatorname{sinc}(t\epsilon)$.\\ 
\hspace*{6mm}
In order to show that  $|t|$ is minimal assume that $0\le\delta <|t| $ is an exponent for  
$\operatorname{sinc}(t\epsilon)$. Let $\delta<\delta'<|t|$. Then obviously $|\sin\big(t\epsilon(z)\big)|\le C\operatorname{e}^{\delta'|z|}$, $z\in\C^3$ for some finite constant $C$. Let $w\in\C$. Choose $\zeta\in\C $ with $\zeta^2=w^2-m^2$. Then $w\in\{\pm\epsilon(0,0,\zeta)\}$ and $|\zeta|\le |w|+m$. Hence $|\sin(tw)|\le C\operatorname{e}^{\delta'|\zeta|}\le C'\operatorname{e}^{\delta'|w|}$ with $C':=C\operatorname{e}^{\delta'm}$.
Therefore also $|\cos(tw)|=|\sin(tw+\frac{\pi}{2})|\le C'\operatorname{e}^{\frac{\pi\delta'}{2|t|}}\operatorname{e}^{\delta'|w|}$, whence finally
 $|\operatorname{e}^{tw}|\le C''\operatorname{e}^{\delta'|w|}$, $w\in\C$ for some finite constant $C''$. This implies  the contradiction $\operatorname{e}^{(|t|-\delta')r}\le C''$ for all $r>0$.\\
\hspace*{6mm} 
 We turn to the P-indicator of $\operatorname{sinc}(t\epsilon)$. Assume at once $\lambda\ne 0$. Then $\epsilon(p+\operatorname{i}\lambda r)=\big(\mu^2+(\frac{p\lambda}{|\lambda|}+\operatorname{i}|\lambda|r)^2\big)^{1/2}$ with $\mu^2:=m^2+p^2-\big(\frac{p\lambda}{|ß\lambda |}\big)^2\ge 0$ independent of $r$. Hence by 
(\ref{EFSINC}) there are finite constants $A_t>0$, $B_t$ independent of $r$ that such $A_t\big|\frac{p\lambda}{|\lambda|}+\operatorname{i} |\lambda| \,r\big|^{-1}\operatorname{e}^{|t||\lambda|r}\le \operatorname{sinc}\big(t\epsilon(p+\operatorname{i}\lambda r)\big)\le B_t\big|\frac{p\lambda}{|\lambda|}+\operatorname{i} |\lambda| \,r\big|^{-1}\operatorname{e}^{|t||\lambda|r}
$, whence the assertion.\qed\\

\textbf{Proof of\, (\ref{GGETIIED})\,Theorem.} Start from (\ref{DRDTE}) $\phi=\operatorname{i}t\operatorname{sinc}(t\epsilon)\,\chi$. Put here $\phi_{kl}:=(\varphi_t)_k\varphi_l$, $\chi_{kl}:=(h\varphi)_k\varphi_l$, whence $\phi=\phi_{kl}-\phi_{lk}$ and $\chi=\chi_{kl}-\chi_{lk}$.\\
\hspace*{6mm}
As $\varphi_l|_{\R^3} \in L^2$ and $(\varphi_t)_k|_{\R^3}$ is bounded,  $\phi_{kl}|_{\R^3}\in L^2$ so that 
(\ref{TPP}) applies to  $\phi_{kl}$. Let $\theta:=\mathcal{F}^{-1}\phi|_{\R^3}$, $\theta_{kl}:=\mathcal{F}^{-1}\phi_{kl}|_{\R^3}$. Obviously, $e(\theta)\ge \min\{e(\theta_{kl}),e(\theta_{lk})\}$.
Using (\ref{REH}) one gets $e(\theta_{kl})=-H_{C(\theta_{kl})}(-e)=-h_{\phi_{kl}}(-e)=-h_{(\varphi_t)_k}(-e)-h_{\varphi_l}(-e)=-H_{C((\psi_t)_k)}(-e)-H_{C(\psi_l)}(-e)=e((\psi_t)_k)+e(\psi_l)\ge e(\psi_t)+e(\psi)$. It follows $e(\theta)\ge e(\psi_t)+e(\psi)$.\\
\hspace*{6mm}
We turn to the right hand side\, $\operatorname{i}t\operatorname{sinc}(t\epsilon)\,\chi$ of (\ref{DRDTE}).  Recall  $\operatorname{i}t\operatorname{sinc}(t\epsilon)\,\chi_{kl}=\phi_{kl}-\cos(t\epsilon)\varphi_k\varphi_l$ by (\ref{TED}). Note that $\cos(t\epsilon)\varphi_k|_{\R^3}$ is bounded. Hence $\operatorname{sinc}(t\epsilon)\,\chi_{kl}|_{\R^3}\in L^2$. However, $\chi_{kl}|_{\R^3}$ need not be square-integrable. Therefore  we consider instead $\chi'_{kl}:=s_\delta \chi_{kl}$ with $s_\delta:=\operatorname{sinc}(\delta\epsilon)$  for $\delta>0$. Then $\phi'=\operatorname{i}t\operatorname{sinc}(t\epsilon)\,\chi'$ for $\phi':=s_\delta\phi$ holds. As $h_{s_\delta}(e)=\delta$  by (\ref{PICESCE}), the analogous computation for 
$\theta':=\mathcal{F}^{-1}\phi'|_{\R^3}$ in place of $\theta$ yields $e(\theta')\ge -\delta+e(\psi_t)+e(\psi)$.
Moreover, (\ref{TPP}) applies to $\chi'$. Let $\xi':=\mathcal{F}^{-1}\chi'|_{\R^3}$. Then again, in the same way $e(\theta')=-|t|+e(\xi')$ follows.\\
\hspace*{6mm}
Next we examine  $-\overline{e}(\xi')$. Obviously $-\overline{e}(\xi')\le\max\{-\overline{e}(\xi'_{kl}),-\overline{e}(\xi'_{lk})\}$. By (\ref{REH}) and (\ref{TPP}) one has $-\overline{e}(\xi'_{kl})=H_{C(\xi'_{kl})}(e)=h_{\chi'_{kl}}(e)=h_{s_\delta(h\varphi)_k}(e)+h_{\varphi_l}(e)$ as $s_\delta(h\varphi)_k$ is exponentially bounded. Note  $|(h\varphi)_k(z)|\le q(z)\max_m|\varphi_m(z)|$ with
$q(z)^2:=4\sum_{m=1}^4|h(z)_{km}|^2$, where $h(z)_{km}$ is linear. Therefore $h_{s_\delta(h\varphi)_k}(e,x)=\varlimsup_{r\to\infty}\frac{1}{r}\big\{\ln|s_\delta(x+\operatorname{i}e\,r))|+ \ln|(h\varphi)_k(x+\operatorname{i}e\,r)|\big\}=\delta+\varlimsup_{r\to\infty}\frac{1}{r} \ln|(h\varphi)_k(x+\operatorname{i}e\,r)|\}$ (by (\ref{PICESCE})) $\le \delta+\varlimsup_{r\to\infty}\frac{1}{r}\big\{\ln|q(x+\operatorname{i}e\,r)|+\ln(\max_m|\varphi_m(x+\operatorname{i}e\,r)|)\big\}=\delta+0+\max_m\varlimsup_{r\to\infty}\frac{1}{r}\ln|\varphi_m(x+\operatorname{i}e\,r)|=\delta+\max_m h_{\varphi_m}(e,x)$. Furthermore, $\max_m h_{\varphi_m}(e)=\max_mH_{C(\psi_m)}(e)=\max_m\{-\overline{e}(\psi_m)\}=-\overline{e}(\psi)$. Also $h_{\varphi_l}(e)\le -\overline{e}(\psi)$. It follows  $-\overline{e}(\xi')\le \delta-2\,\overline{e}(\psi)$.\\
\hspace*{6mm}  
 Now, using $e(\xi')<- \overline{e}(\xi')$, one has the chain of inequalities $ -\delta+e(\psi_t)+e(\psi)\le e(\theta')=-|t|+e(\xi')<-|t| -\overline{e}(\xi')\le -|t|+ \delta-2\,\overline{e}(\psi)$ for $\delta>0$. The limit $\delta\to 0$ yields the final result $ e(\psi_t)+e(\psi)\le -|t|+-2\,\overline{e}(\psi)$. It remains to note that if $h\varphi$ is bounded on $\R^3$ one has $\chi_{kl}|_{\R^3}\in L^2$ so that $e(\xi)<- \overline{e}(\xi)$, and the chain holds even for $\delta=0$.
\qed\\

The next two lemmas serve for the proof of (\ref{SETIIED}).  

\begin{Lem}\label{SMF} Let $\psi\ne0$ be a Dirac wave function  localized in a bounded region. Then 
\begin{equation*}
 \min\{e(\psi_t),e(\psi_{-t})\}= e(\psi) -|t| 
\end{equation*}
holds for every direction $e$  and  all times $t\in\R$.
\end{Lem}\\
 {\it Proof.} 
 By causality $e(\psi_t)\ge e(\psi) -|t|$ for all $t$, whence $\min\{e(\psi_t),e(\psi_{-t})\} \ge e(\psi)-|t|$. \\
\hspace*{6mm} 
 We prove now  the reverse inequality. Recall   $\phi= 2\cos(t\epsilon)\,\varphi$ for $\phi:=\varphi_t+\varphi_{-t}$ from (\ref{STED}).  Let $\theta:=\mathcal{F}^{-1}\phi|_{\R^3}$. Theorem (\ref{TPP}) applies to the components of $\varphi$ and, due to (\ref{PICESCE}), also to those of $\cos(t\epsilon)\,\varphi$.
Hence, using (\ref{REH}) and by (\ref{PICESCE}), $e(\theta_l)=-H_{C(\theta_l)}(-e)=-h_{\cos(t\epsilon)\,\varphi_l}(-e)=-h_{\cos(t\epsilon)}(-e)-h_{\varphi_l}(-e)=-|t|-H_{C(\psi_l)}(-e)=-|t|+e(\psi_l)$. Therefore $e(\theta)=\min_le(\theta_l)=-|t|+\min_le(\psi_l)=-|t|+e(\psi)$.\\
\hspace*{6mm}
It remains to show $\alpha:=\min\{e(\psi_t),e(\psi_{-t})\}\le e(\psi_t+\psi_{-t})$. Put $\chi_\alpha:= 1_{\{x\in\R^3: x\operatorname{e}\le\alpha\}}$. Then $\chi_\alpha\psi_t=0$ and $\chi_\alpha\psi_{-t}=0$. Hence  $\chi_\alpha(\psi_t+\psi_{-t})=0$, whence the claim. \qed\\

\begin{Lem}\label{CDFE} Let $\psi$ be a Dirac wavefunction.  
Then $\R\to\R$, $t\mapsto e(\psi_t)$ is continuous.
\end{Lem}\\
{\it Proof.} Let $t,t_0\in\R$. By causality, $e(\psi_t)\ge e(\psi_{t_0})-|t-t_0|$. This implies $\varliminf_{t\to t_0} e(\psi_t)\ge e(\psi_{t_0})$. Furthermore, for $\chi_t:= 1_{\{x:x\,e\le e(\psi_t)\}}$ one has $0=\chi_t\psi_t=\chi_t\psi_{t_0} +\chi_t(\psi_t-\psi_{t_0})$, whence $\lim_{t\to t_0}\chi_t\psi_{t_0}=0$ as $\psi_t\to \psi_{t_0}$. This implies $\varlimsup_{t\to t_0}e(\psi_t)\le e(\psi_{t_0})$. Thus continuity of $t\to e(\psi_t)$  at $t_0$ holds.\qed\\

\textbf{Proof of\, (\ref{SETIIED})\,Theorem.} Since $t\to e(\psi_t)$ is  continuous by  (\ref{CDFE}) and bounded   above by (\ref{GGETIIED}) there is $t_e\in\R$ with $e(\psi_{t_e})=\sup_{t\in\R}e(\psi_t)$. Fix $t>0$.\\
\hspace*{6mm}
Now we apply (\ref{SMF})
to $\psi':=\psi_{t_e-t/2}$. Then $\min\{e(\psi'_{t'}),e(\psi'_{-t'})\} =e(\psi') -|t'|$ for all $t'\in\R$. As $e(\psi'_{t/2})=e(\psi_{t_e})\ge e(\psi_{t_e-t})=e(\psi'_{-t/2})$ it follows $e(\psi_{t_e-t})=e(\psi_{t_e-t/2})-t/2$. For $t/2$ in place of $t$ this reads  $e(\psi_{t_e-t/2})=e(\psi_{t_e-t/4})-t/4$. Hence $e(\psi_{t_e-t})=e(\psi_{t_e-t/4})-t/2-t/4$. From this one obtains in the same way $e(\psi_{t_e-t})=e(\psi_{t_e-t/8})-t/2-t/4-t/8$ and  finally $e(\psi_{t_e-t})=e(\psi_{t_e-t/2^n})-\sum_{k=1}^n t/2^k$ after  $n$ steps. Then by continuity (\ref{CDFE}) the limit $n\to \infty$ yields  $e(\psi_{t_e-t})=e(\psi_{t_e})-t$. --- 
Analogously, applying  (\ref{SMF}) to $\psi':=\psi_{t_e+t/2}$, one obtains  $e(\psi_{t_e+t})=e(\psi_{t_e})-t$.\\
\hspace*{6mm}
Thus  $e(\psi_t)= e(\psi_{t_e})-|t-t_e|$ holds for all $t\in\R$. In particular $e(\psi)= e(\psi_{t_e})-|t_e|$, whence the formula.
 Uniqueness of $t_e$ is obvious as $t\to e(\psi_t)$ has just one maximum at $t=t_e$.\qed\\

\textbf{Proof of\, (\ref{CSETIIED})\,Corollary.} (a) By (\ref{SETIIED})  and  (\ref{GGETIIED}) one has $e(\psi_t)=e(\psi) +|t_e|-|t-t_e|\le -2\overline{e}(\psi)-e(\psi)-|t|$. For $t=t_e$ this yields $|t_e|\le -\overline{e}(\psi)-e(\psi)$ and consequently $e(\psi_t)\le -\overline{e}(\psi)-|t-t_e|$. 
\\
\hspace*{6mm}
(b) Let $s,t\in\R$ and consider $e(\psi_{t+s})$. One the one hand, by (\ref{SETIIED}), $e(\psi_{t+s})=e(\psi)+|t_e|-|t+s-t_e|$. On the other hand, first using (\ref{GGETIIED}) and then applying (\ref{SETIIED}), one has $e(\psi_{t+s})\le -2\overline{e}(\psi_t)-e(\psi_t)-|s|= 
-2\big(\overline{e}(\psi)+|t_{\overline{e}}|-|t-t_{\overline{e}}|\big)-e(\psi)-|t_e|+|t-t_e|-|s|$. Hence 
$-2\big(\overline{e}(\psi)+e(\psi)\big)\ge 2|t_e|+2|t_{\overline{e}}|-2|t-t_{\overline{e}}|-|t-t_e| +|s|-|t+s-t_e|$. For $s=t_e-t$ this yields $-\overline{e}(\psi)-e(\psi)\ge |t_e|+|t_{\overline{e}}|-|t-t_{\overline{e}}|$. Then  $|t_e|+|t_{\overline{e}}|\le -\overline{e}(\psi)-e(\psi)$ follows for $t=t_e$. \\
\hspace*{6mm}
(c) By (\ref{EEP}), $0<-\overline{e}(\psi_t)-e(\psi_t)$ for all $t$.   Hence (\ref{SETIIED}) yields $|t_{\overline{e}}|+|t_e|-|t-t_{\overline{e}}|-|t-t_e|<-\overline{e}(\psi)-e(\psi)$.
This implies (c).\\
\hspace*{6mm}
(d) Let $\tau\in\R$, $\psi':=\psi_\tau$, and $t'_e:=t_e(\psi')$. Then $e(\psi'_t)= e(\psi')+|t'_e|-|t-t'_e|$ and $e(\psi')=e(\psi)+|t_e|-|\tau-t_e|$. As $\psi'_t=\psi_{t+\tau}$ also $e(\psi'_t)=e(\psi)+|t_e|-|t+\tau -t_e|$ holds. Therefore $|t+\tau-t_e|-|t-t'_e|=|\tau-t_e|-|t'_e|$ for all $t$, whence $t'_e=t_e-\tau$.\\
\hspace*{6mm}
(e)  follows from the last part of  (\ref{GGETIIED}).
 \qed\\

\textbf{Proof of\, (\ref{LTSIE})\,Corollary.} By (\ref{SETIIED}) one has $e(\psi_t)=e(\psi_{t_e})-|t-t_e|$.  For $t\ge 2R\ge |t_e|$ by (\ref{CSETIIED})(b) it follows $ e(\psi_t)=e(\psi_{t_e})+t_e-t$ and in particular $ e(\psi_{2R})=e(\psi_{t_e})+t_e-2R$, whence $ e(\psi_t)=e(\psi_{2R})+2R-t $. Similarly, for $t\le -2R$ one has $ e(\psi_t)=e(\psi_{t_e})-t_e+t$ and in particular $ e(\psi_{-2R})=e(\psi_{t_e})-t_e-2R$, whence $ e(\psi_t)=e(\psi_{-2R})+2R+t $.\qed\\

The space translations $b\in\R^3$ act on the Dirac wavefunctions $\psi$ by  $\big(W(b)\psi\big)(x):=\psi(x-b)$.
For the following construction we use the easily verifiable formulae 
\begin{equation}\label{TCTTE}
e\big(W(\lambda e)\psi\big)=e(\psi)+\lambda, \quad t_e\big(W(\lambda e)\psi\big)=t_e(\psi)
\end{equation} 
for all directions $e$ and $\lambda\in\R$.  
\\

\textbf{Proof of\, (\ref{NISE})\,Lemma.}  Let $\tau\in\R\setminus\{0\}$ and $\delta>0$. 
Let $\psi^{(1)}\ne 0$ be any  bounded localized wavefunction.  
 Set $\psi^{(2)}:=W(\delta e)\psi^{(1)}_\tau$ and put
 $$\psi:=\psi^{(1)}+\psi^{(2)}$$
  In the following we  express the characteristic dates  $e(\psi),t_e$,  $\overline{e}(\psi), t_{\overline{e}}$ referring to $\psi$ by the input  dates 
$e(\psi^{(1)}),t^{(1)}_e$,  $\overline{e}(\psi^{(1)}), t^{(1)}_{\overline{e}}$ and the parameters $\tau, \delta$.\\
\hspace*{6mm} 
By  (\ref{SETIIED}), (\ref{CSETIIED}) and (\ref{TCTTE}) one has
$t^{(2)}_e=t^{(1)}_e-\tau$ and $t^{(2)}_{\overline{e}}=t^{(1)}_{\overline{e}}-\tau$,
and  $e(\psi^{(1)}_t)=e(\psi^{(1)})+|t^{(1)}_e|-|t-t^{(1)}_e|$,  $e(\psi^{(2)}_t)=e(\psi^{(1)}_{t+\tau }) +\delta=e(\psi^{(1)})+|t^{(1)}_e|-|t+\tau-t^{(1)}_e|+\delta$ and similarly $\overline{e}(\psi^{(1)}_t)=\overline{e}(\psi^{(1)})+|t^{(1)}_{\overline{e}}|-|t-t^{(1)}_{\overline{e}}|$, $\overline{e}(\psi^{(2)}_t)=\overline{e}(\psi^{(1)})+|t^{(1)}_{\overline{e}}|-|t+\tau-t^{(1)}_{\overline{e}}|-\delta$.\\
 \hspace*{6mm} 
 Obviously $e(\psi_t)=\min\{e(\psi^{(1)}_t),e(\psi^{(2)}_t)\}$ and  $\overline{e}(\psi_t)=\min\{\overline{e}(\psi^{(1)}_t),\overline{e}(\psi^{(2)}_t)\}$. Hence $t_e$ and $t_{\overline{e}}$ are determined by  (\ref{SETIIED}). Write $e(\psi^{(2)}_t)-e(\psi^{(1)}_t) =
 d(t-t_e^{(1)})$ with $d(x):=|x|-|x+\tau|+\delta$  and $\overline{e}(\psi^{(2)}_t)-\overline{e}(\psi^{(1)}_t) =
 \overline{d}(t-t^{(1)}_{\overline{e}})$ with $\overline{d}(x):=|x|-|x+\tau|-\delta$. Note
 \begin{equation*}
 |\tau|\le\delta\, \Leftrightarrow\,  d(t-t_e^{(1)})\ge 0 \; \forall t \, \Leftrightarrow\,  \overline{d}(t-t^{(1)}_{\overline{e}})\le 0 \; \forall t\tag{$\star$}
\end{equation*}
Indeed, $ d(t-t_e^{(1)})\ge 0$ is equivalent to $ |\tau|\le\delta$ as $d$ takes
 its minimum $-|\tau|+\delta$ at 
 $x=0$. Similarly, $\overline{d}$ takes its maximum $|\tau|-\delta$ at $x=-\tau$.\\
\hspace*{6mm}
Now consider  the case $|\tau|\le\delta$. By ($\star$) one has $e(\psi_t)=e(\psi^{(1)}_t)$, $\overline{e}(\psi_t)= \overline{e}(\psi^{(2)}_t)$,  whence $t_e=t^{(1)}_e$ and $t_{\overline{e}}=t^{(2)}_{\overline{e}}=t^{(1)}_{\overline{e}}-\tau$.  
So one obtains the given value of
$ t_{\overline{e}}-t_e$  by choosing $\tau =(t^{(1)}_{\overline{e}}-t^{(1)}_e)-(t_{\overline{e}}-t_e)$. By a subsequent time translation, according to (\ref{CSETIIED})(d) one gets the prescribed values of $t_e$ and $t_{\overline{e}}$. \qed

 The construction  in (\ref{NISE}) for $\tau:=t^{(1)}_{\overline{e}}-t^{(1)}_e$, $\delta:=|\tau|$ with subsequent time translation by $t_e=t_{\overline{e}}$ yields a Dirac wavefunction  $\psi$ satisfying
 \begin{equation}\label{TETEOEZ} 
t_e=t_{\overline{e}}=0 \textrm{\, and } -\overline{e}(\psi)-e(\psi)= -\overline{e}(\psi^{(1)})-e(\psi^{(1)}) +
|t^{(1)}_{\overline{e}}-t^{(1)}_e |-|t^{(1)}_{\overline{e}}|- | t^{(1)}_e |
\end{equation}
The width of the carrier in direction $e$ in not increased since $|t^{(1)}_{\overline{e}}-t^{(1)}_e |-|t^{(1)}_{\overline{e}}|- | t^{(1)}_e |\le 0$.\\

\textbf{Proof of\, (\ref{DSABTP})\,Lemma.} Due to (\ref{TCTTE}) it is no restriction to assume $a=-b$. Let $0<\rho<b$. By (\ref{TETEOEZ}) there is  a Dirac wavefunction $\eta$ localized in a bounded region contained in $\{-\rho\le xe\le \rho\}$ with $t_e(\eta)=t_{\overline{e}}(\eta)=0$. Let $\varsigma$ denote the sign of $\tau$. Then, by causality, $\psi:=\eta_{\varsigma(-b+\rho)}$ is localized in $\{-b\le xe\le  b\}$. Moreover, $t_e=t_{\overline{e}}= \varsigma(b-\rho)$ holds by (\ref{CSETIIED})(d). The assertion follows for $\rho:=b-|\tau|$.\qed\\

The main mathematical tool for the proofs of the claims in sec.\,\ref{LTBPL}  is an application of the non-stationary phase method as shown in \cite[Theorem 1.8.]{T92} estimating (\ref{ENSPM})  for large  $|x|+|t|$. The  result in (\ref{PRBG}), according to which the spatial probability in $B_{r}$ tends to zero, essentially 
 is a corollary to  \cite[Corollary 1.9.]{T92}. 
 Rather analogously we prove in (\ref{VPOBT})  the fact  that  asymptotically  the spatial probability vanishes outside $B_{|t|}$.\\
\hspace*{6mm}
In the following  the  obvious reduction to scalar-valued wavefunctions   is used. Let $\psi\in L^2(\R^3,\C^4)$ be a Dirac wavefunction and let $\varphi=\mathcal{F}\psi$ be its momentum representation. Regarding the time translation one has $\varphi_t=\operatorname{e}^{\operatorname{i}th}\varphi$, i.e., $\varphi_t(p)=\operatorname{e}^{\operatorname{i}th(p)}\varphi(p)$ for  $p\in\R^3$.  Let $\eta\in\{+,-\}$ and note that $\pi^\eta(p)=\frac{1}{2}(I+\frac{\eta}{\epsilon(p)}h(p))$ with $\epsilon(p)=\sqrt{|p|^2+m^2}$ is the projection in $\C^4$ onto the $2$-dimensional eigenspace of $h(p)$ with eigenvalue $\eta\, \epsilon(p)$.  Then  $\varphi^\eta:=\pi^\eta\varphi$
is the projection  of $\varphi$ onto the positive, respectively negative, energy eigenspace. Analogously 
 $(\varphi_t)^\eta:=\pi^\eta\varphi_t$. Note that $(\varphi_t)^\eta=(\varphi^\eta)_t=\operatorname{e}^{\operatorname{i}t\eta \epsilon}\varphi^\eta$, as $\operatorname{e}^{\operatorname{i}th}$ and $\pi^\eta$ commute. One concludes $(\psi_t)_l=\sum_\eta  (\psi^\eta_t)_l$ with    $ (\psi^\eta_t)_l:= \big(\mathcal{F}^{-1}\varphi^\eta_t\big)_l=
  \mathcal{F}^{-1}\big(\operatorname{e}^{\operatorname{i}t\eta \epsilon}(\varphi^\eta)_l\big)$ for the $l$-th component of $\psi_t$, $l=1,\dots,4$. If $\varphi$  is also  integrable, then so is $\varphi^\eta$ and for each $l$ one has
 \begin{equation}\label{ENSPM}
 (\psi^\eta_t(x))_l=(2\pi)^{-3/2}\int \operatorname{e}^{\operatorname{i}(px+t\eta\epsilon(p))}(\varphi^\eta(p))_l\,\operatorname{d}^3p
 \end{equation}

\textbf{Proof of (\ref{PRBG}) Theorem.} Recall $\varphi=\mathcal{F}\psi$ and choose $\varphi'\in\mathcal{C}^\infty_c(\R^3\setminus \{0\},\C^4)$ with $\norm{\varphi-\varphi'}\le\varepsilon/2$. Hence $\norm{\psi-\psi'}\le\varepsilon/2$ for $\psi'=\mathcal{F}^{-1}\varphi'$. Choose $0<v<\inf\{\frac{|p|}{\epsilon(p)}:p\in \operatorname{supp}(\varphi')\}$.  Let $\chi_t := 1_{B_{v|t|}}$. Now, according to \cite[Corollary 1.9.]{T92},  there is a constant $C_1$ such that $\norm{\chi_t\psi'_t}\le
C_1(1+|t|)^{-1}$ for all $t$. Let $\tau:=2C_1/\varepsilon$. Then $\norm{\chi_t\psi_t}\le \norm{\chi_t(\psi_t-\psi'_t)} + \norm{\chi_t\psi'_t}  \le$ $\norm{\psi-\psi'} +\,     C_1(1+|t|)^{-1} \le\varepsilon$ for $|t|\ge\tau$. --- Now fix $r>0$. Then for $|t|\ge \max\{\tau, \frac{r}{v}\}$ one has $\norm{1_{B_r}\psi_t}\le \norm{\chi_t\psi_t}\le \epsilon$.
\qed\\

\textbf{Proof of (\ref{VPOBT}) Theorem.}  Suppose first $\varphi:=\mathcal{F}\psi\in \mathcal{C}^\infty_c(\R^3,\C^4)$. Let $K:=\operatorname{supp}(\varphi)$. Set $\gamma:=\max\{\frac{|p|}{\epsilon(p)}:p\in K\}$. Clearly $0<\gamma<1$. For the estimation of the integral in (\ref{ENSPM})  consider $\phi^\eta(p):= (|x|+|t|)^{-1}\big(px-t\eta\epsilon(p)\big)$. Then $\nabla \phi^\eta(p)=  (|x|+|t|)^{-1}\big(x-\frac{t\eta}{\epsilon(p)}p\big)$ and $|\nabla \phi^\eta(p)|\ge  (|x|+|t|)^{-1}\big(|x|-|t|\frac{|p|}{\epsilon(p)}\big)\ge \frac{|x|-\gamma|t|}{|x|+|t|}$ for $p\in K$. Now suppose $|x|\ge|t|$. Then $|\nabla \phi^\eta(p)|\ge  \frac{|x|-\gamma|x|}{|x|+|x|}=\frac{1-\gamma}{2}>0$. This implies  (cf.\,\cite[(1.209)]{T92}) for $\eta\in\{+,-\}$, $l=1,\dots,4$, and for every $N>0$ that there is a finite constant $A_N$ with
$$\big|\big(\psi^\eta_t(x)\big)_l\big|\le A_N(1+|x|+|t|)^{-N}\textrm{ \;if\; } |x|\ge |t|$$
Put $\chi_t:=1_{\R^3\setminus B_{|t|}}$. Then $\norm{\chi_t\psi_t}\le \sum_\eta \norm{\chi_t\psi^\eta_t}$ and  by the above estimation 
$\norm{\chi_t\psi^\eta_t}^2=\int_{\R^3\setminus B_{|t|}}\norm{\psi^\eta_t(x)}^2\operatorname{d}x^3=\sum_l\int_{\R^3\setminus B_{|t|}}\big|\big(\psi^\eta_t(x)\big)_l\big|^2\operatorname{d}x^3\le 16\pi A_N^2\int_{|t|}^\infty (1+r+|t|)^{-2N} r^2\operatorname{d}r\le 16\pi A_N^2\int_{|t|}^\infty (1+r)^{-2N+2}\operatorname{d}r=\frac{16\pi}{2N-3}A_N^2(1+|t|)^{-2N+3}$ if $N>\frac{3}{2}$. Hence $\norm{\chi_t\psi_t}\le C_N(1+|t|)^{-N}$ for $N>0$ and  $C_N:=(32\pi/N)^{\frac{1}{2}}A_{N+\frac{3}{2}}$. \\
\hspace*{6mm}
Now consider a general Dirac wavefunction $\psi$. Let $\varepsilon>0$. Set $\varphi:=\mathcal{F}\psi$ and choose $\varphi'\in\mathcal{C}^\infty_c(\R^3,\C^4)$ with $\norm{\varphi-\varphi'}\le\varepsilon/2$. Hence $\norm{\psi-\psi'}\le\varepsilon/2$ for $\psi':=\mathcal{F}^{-1}\varphi'$.  By the foregoing result  there is a constant $C_1$ such that $\norm{\chi_t\psi'_t}\le
C_1(1+|t|)^{-1}$ for all $t$. Let $\tau:=2C_1/\varepsilon$. Then $\norm{\chi_t\psi_t} \le \norm{\chi_t(\psi-\psi')_t}  +  \norm{\chi_t\psi'_t} \le$ $\norm{\psi-\psi'}+C_1(1+|t|)^{-1}\le\varepsilon $ for $|t|\ge\tau$.\qed\\


\end{document}